\documentclass[twocolumn, aps, prl]{revtex4-1}
\usepackage{graphicx,bm,amssymb,amsmath,color}

\newcommand\smref{Supplementary Material}

\begin{document}

\title{Minimal design of the elephant trunk as an active filament}  
\author{Bartosz Kaczmarski, Sophie Leanza, Renee Zhao,}
\affiliation{Department of Mechanical Engineering,
Stanford University, Stanford, CA-94305, United States}
\author{Derek E.\ Moulton}
\affiliation{Mathematical Institute, University of Oxford, Oxford, UK}
\author{Ellen Kuhl}
\affiliation{Department of Mechanical Engineering,
Stanford University, Stanford, CA-94305, United States}
\author{Alain Goriely}
\affiliation{Mathematical Institute, University of Oxford, Oxford, UK}

\begin{abstract}
\noindent One of the key problems in active materials is the control of shape through actuation. 
A fascinating example of such control is the elephant trunk, a long, muscular, and extremely dexterous organ with multiple vital functions. 
The elephant trunk is an object of fascination for biologists, physicists, and children alike.
Its versatility relies on the intricate interplay of multiple unique physical mechanisms and biological design principles. 
Here we explore these principles using
the theory of active filaments and build, theoretically, computationally, and experimentally, a minimal model that explains and 
accomplishes some of the spectacular features of the elephant trunk.
\end{abstract}

\maketitle

\textit{Introduction.}---In nature and engineering, one of the main tasks of active filamentary structures is to explore their surroundings. 
The simplest problem for the activation of slender structures is curvature generation, a well-understood phenomenon studied by Timoshenko a century ago in the context of bi-metallic strips \cite{ti25}. In Timoshenko's initial setting, curvature changes only generate planar shapes. A more involved problem is the generation of torsion which, through curvature coupling, allows for arbitrary three-dimensional shapes as found in plants \cite{moulton2020multiscale}, but also in the octopus arms and elephant trunks \cite{dagenais2021elephants}. Here we take inspiration from the elephant trunk as a design paradigm to couple curvature, twist, and torsion, and create a minimal activation model using three uniform actuators.

\begin{figure}
 \centering
\includegraphics[width=0.99\columnwidth]{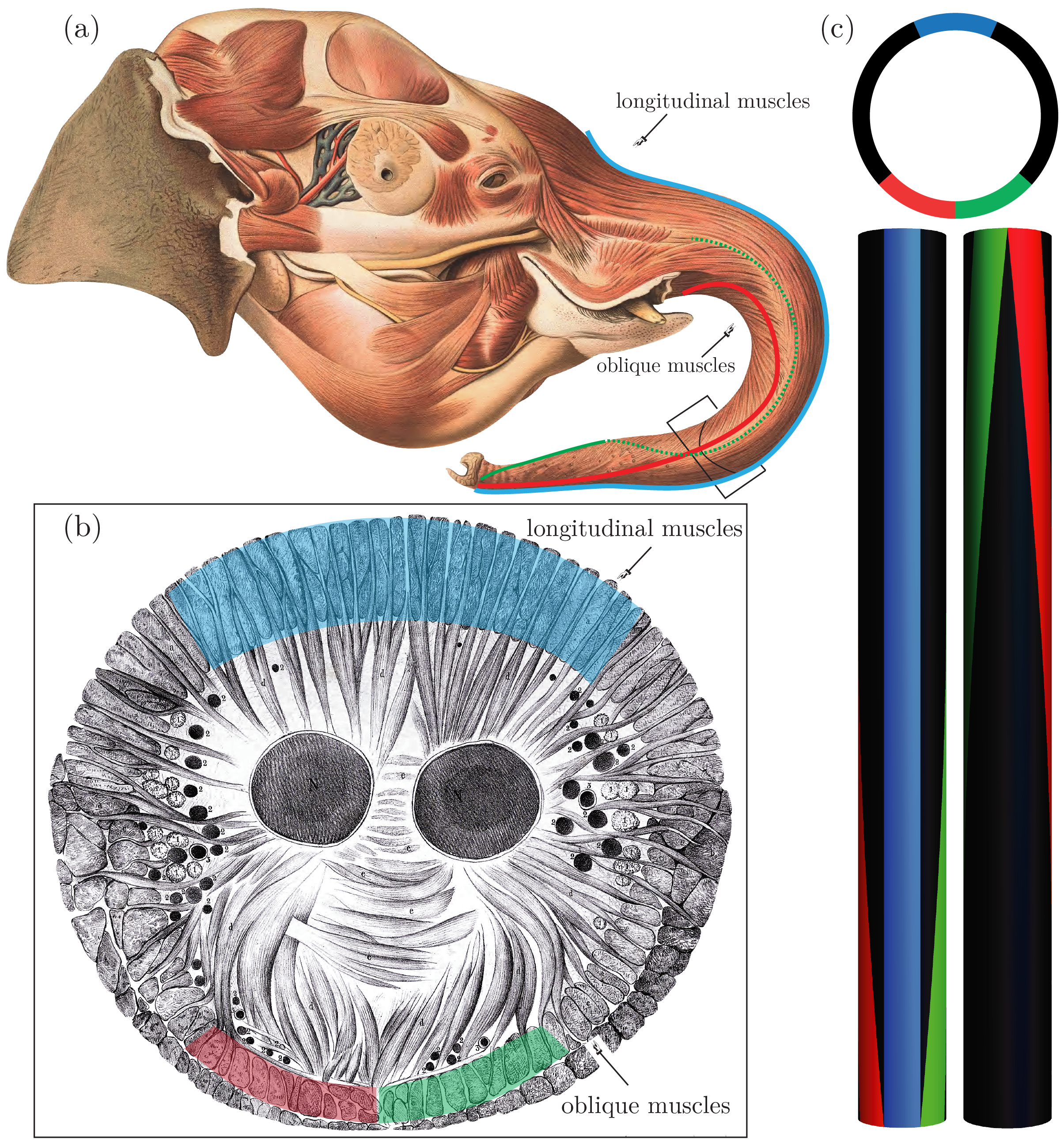}
\caption{{\bf{Muscular architecture within the elephant trunk.}} (a) Skinned elephant trunk with longitudinal muscles on the dorsal side and oblique muscles wrapping helically along the length on the ventral side \cite{boas1908elephant}. (b) Cross ection of a trunk by Cuvier (1850)  with longitudinal (blue), oblique (red, green) and radial muscles around the two nasal cavities \cite{cuvier1850anatomie}. (c) Minimal model of a trunk with three muscular bundles.}\label{fig1}
\end{figure}

The elephant trunk is an elongated, muscular, and highly flexible proboscis made of specialized muscles, fascia, and skin; it represents a unique adaptation of the upper lip and the nose, see Fig.~\ref{fig1}. 
It is made of  17 major muscle groups, eight on each side and one for the nasal cavity, with more than 90,000 muscle fascicles  controlled by 50-60,000 facial nucleus neurons, which together give it its exceptional strength and versatility of movement \cite{kaufmann2022elephant,purkart2022trigeminal,longren2023dense}.

Unsurprisingly, the biomechanics of the elephant trunk have attracted the attention of many scientists \cite{wu2018elephant,schulz2022skin} with the ultimate goal to understand and replicate some of its functions \cite{guan2020novel}. Yet, some of the basic principles of the trunk have been overlooked. Here, in contrast to most previous studies that use a great number of segmented actuators to replicate function \cite{trivedi2008soft,zhang2023preprogrammable}, we aim to unravel important properties of the trunk by considering an ideal trunk with a {\it{minimum number}} of actuators that still reproduces some of its key functions. Towards this goal, we show that simple physical models of active materials can help improve our understanding of these spectacular structures. 
Inspired by the muscular architecture of the trunk, we propose a minimal model of three actuators and investigate its control capabilities. In particular, we  study the design space of our idealized trunk and identify the fundamental trade-off between some desirable properties. Then, we validate our model predictions experimentally using liquid crystal elastomer actuators. 

\begin{figure}
\centering
\includegraphics[width=1.0\columnwidth]{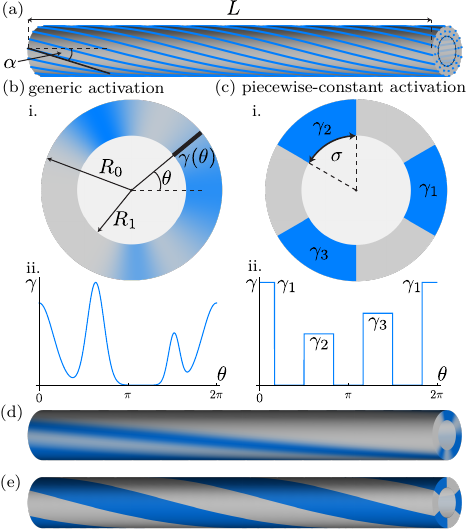}
\caption{{\bf{Active filament model.}} (a) Filament with embedded helically-arranged fibers. (b) and (d) Realistic continuous muscle activation $\gamma(\theta)$. (c) and (e) Idealized discrete piecewise constant activation with $N=3$ actuators.}
\label{fig2}
\end{figure}

\textit{Model set-up.}---Despite its massive size, the elephant trunk can be modeled as a slender biological filament that deforms by muscular activation of longitudinal, radial, and helical muscles (Fig.~\ref{fig1}). 
To model the deformation induced by local fiber contraction, we adopt the active filament theory \cite{kaczmarski2023bayesian,kaczmarski2023simulation,kaczmarski2022active}, which describes a general three-dimensional tubular structure equipped with an arbitrary fiber activation tensor $\mathbf{G}$ dictated by the local orientation of the muscle fibers throughout the continuum. Assuming that initially the trunk is straight, any deformation can be written as $\bm{\chi} = \bm{r} + \sum_{i = 1}^3 \varepsilon e_i\mathbf{d}_i$ \cite{moulton2020morphoelastic}, where $\bm{r}(Z)$ is the centerline of the deformed tubular structure w.r.t.\ the arc length $Z$ of the initial structure, such that $\bm{r}'=\zeta \mathbf{d}_3$, where $\zeta$ is the centerline stretch, $\varepsilon e_i$ are the cross-sectional reactive strains with $\varepsilon$  the slenderness ratio (a characteristic radius divided by the length $L$), and $\{\mathbf{d}_1,\mathbf{d}_2,\mathbf{d}_3\}$ form an orthonormal basis that defines the orientation of each cross-section so that  $\mathbf{d}_i'=\zeta\mathbf{u}\times \mathbf{d}_i$, where $\mathbf{u}=\sum{\mathsf u}_i\mathbf{d}_i $ is the Darboux vector specifying the filament's three curvatures.
To model activation, we decompose the deformation gradient \cite{goriely17}, $\mathbf{F} = \text{Grad}{\bm\chi}=\mathbf{A}\cdot\mathbf{G}$, into an elastic contribution $\mathbf{A}$ and an activation $\mathbf{G}$.
Using a systematic energy minimization procedure, we then obtain the intrinsic curvatures $\hat{{\mathsf u}}_1$, $ \hat{{\mathsf u}}_2$, $\hat{{\mathsf u}}_3$, and the extension $\hat{\zeta}$ of the trunk in the absence of external loads by integrating the contribution of each fiber contraction over the cross-section \cite{kaczmarski2022active}. In the case of a ring of  helical fibers with helical angle $\alpha$ and radii $R_1$ and $R_0$ (see Fig.~\ref{fig2}), the curvatures are:
\begin{align*}
\hat{{\mathsf u}}_{1} (Z)&= -\frac{4}{3R_0^4}A\delta_1\sin\left(\varphi - \frac{Z}{R_0}\tan\alpha\right),\\
\hat{{\mathsf u}}_{2} (Z)&= -\frac{4}{3R_0^4}A\delta_2\cos\left(\varphi - \frac{Z}{R_0}\tan\alpha\right ),\\
\hat{{\mathsf u}}_{3} (Z)&= \frac{2}{R_0^4}\delta_3 a_0,\quad \hat{\zeta} = 1 + \frac{1}{2R_0^2}\delta_0 a_0,
\end{align*}
where $\delta_i$ are functions of the filament geometry, fiber architecture, and mechanical properties (see \smref). For a given periodic muscle activation function $\gamma=\gamma(\theta)$, $A$, $\varphi$, $a_0$ are 
\begin{align*}a_1 &= \frac{1}{\pi}\int_{0}^{2\pi}\!\!\!\!\gamma(\theta)\cos\theta\,\text{d}\theta,\quad
b_1= \frac{1}{\pi}\int_{0}^{2\pi}\!\!\!\!\gamma(\theta)\sin\theta\,\text{d}\theta,\\
a_0&= \frac{1}{\pi}\int_{0}^{2\pi}\!\!\!\!\gamma(\theta)\,\text{d}\theta,\quad A= \sqrt{a_1^2+b_1^2},\quad \tan(\varphi) = -\frac{b_1}{a_1}.
\end{align*}
We restrict our attention to discrete muscle fiber bundles with inputs $\gamma_1,\ldots, \gamma_N$ and angular extents defined by the parameter $\sigma$ as shown in Fig.~\ref{fig2}ce.

\begin{figure*}[t!]
 \centering
\includegraphics[width=1\textwidth]{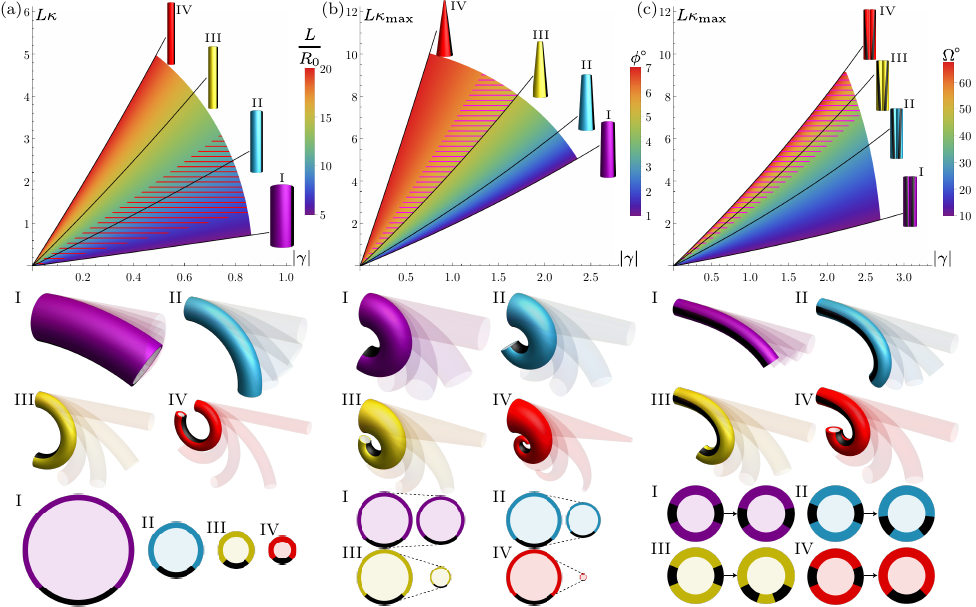}
\caption{{\bf{Design space of the elephant trunk: slenderness, tapering, and fiber helicity.}}
Deformations I, II, III, IV and their respective curvature-activation curves for:
(a) varied slenderness ($S_\text{r} = 5, 10, 15, 20$, $\gamma = -0.6$) with a single longitudinal bundle ($\sigma = 90^{\circ}$); 
(b) varied tapering angle ($\phi = 1^{\circ}, 3^{\circ}, 5^{\circ}, 7^{\circ}$, $\gamma = -2.5$) for the same single bundle; and 
(c) varied helicity ($\Omega \approx {10.0^{\circ}, 29.2^{\circ}, 48.3^{\circ}, 67.5^{\circ}}$, $\gamma = -3$, $\sigma = 45^{\circ}$ in both bundles) for two symmetrically arranged, equally activated helical fiber bundles. 
Dashed regions highlight physiological ranges $S_\text{r}\in[6.7, 11.7]$, and $\phi\in[5^{\circ},6^{\circ}]$ for the elephant trunk, and for $\Omega$ far from the longitudinal case. Young's modulus is constant, Poisson's ratio is $\nu = 1/2$, and $R_0-R_1 = L / 50$ at $Z = 0$.}
\label{fig3}
\end{figure*}

\textit{Design principles.}---We apply the active filament theory to study the main design principles that govern the configuration space of the elephant trunk: slenderness, tapering, and helicity.
First, we consider a single longitudinal muscular bundle. Upon activation, the filament curls to minimize its elastic energy with dimensionless curvature $L\kappa$, extending Timoshenko's result to large deformations. In Fig.~\ref{fig3}a, we show that slenderness has a significant effect on the deformation. Indeed, by varying the ratio $S_\text{r}$ of filament length $L$ to the outer radius $R_0$, the same muscular activation will yield a larger curvature for filaments with a larger slenderness. This is a natural consequence of the scaling of the second moment of area of the filament's cross-section as $\sim 1/S_\text{r}^4$, which dominates the less pronounced effect of the decreased fiber activation area for more slender filaments. Fig.~\ref{fig3}a further shows that, for sufficiently small fiber magnitudes ($|\gamma|<1$), the dimensionless curvature $L\kappa$ is an approximately linear function of the activation in the entirety of the evaluated slenderness range $S_\text{r}\in[5, 20]$. 

Second, the geometry of the elephant trunk is highly tapered, with a larger cross-sectional radius at the proximal end and increasingly smaller radii towards the distal end, effectively inducing a gradually varying slenderness. For  uniform activations in a longitudinal muscle, the local curvature of the trunk increases with the decreasing trunk radius. The larger the decrease in radius the larger the increase in curvature, as demonstrated in Fig.~\ref{fig3}b by the increase in maximal curvature $L\kappa_\text{max}$ with the increasing tapering angle $\phi$. In the physiological range of tapering angles $\phi\in[5^{\circ}, 6^{\circ}]$ (structures III and IV), we observe the typical curling motion that the elephants use to wrap their trunks around objects.

Third, the elephant trunk contains oblique muscle fibers that we model through two opposite helical fibers. In addition to creating curvature, a single helical fiber bundle can generate torsion. Combining the effect of two identical helical fibers  but of opposite handedness cancels the torsional contribution and creates pure bending of the filament \cite{gota13} as shown Fig.~\ref{fig3}c. An increase in the fiber revolution $\Omega$ results in larger maximal curvature and increased variability in the bending curvature (see \smref{} for the definition of the angle $\Omega$). The influence of $\Omega$ on the resulting deformation is similar to the effect of varying the tapering angle $\phi$: the rate of increase of curvature along $Z$ is amplified by increasing either $\Omega$ or $\phi$, which might be responsible for generating the curling deformations commonly observed in elephant trunks motions.

\begin{figure*}[t!]
\centering
\includegraphics[width=1\textwidth]{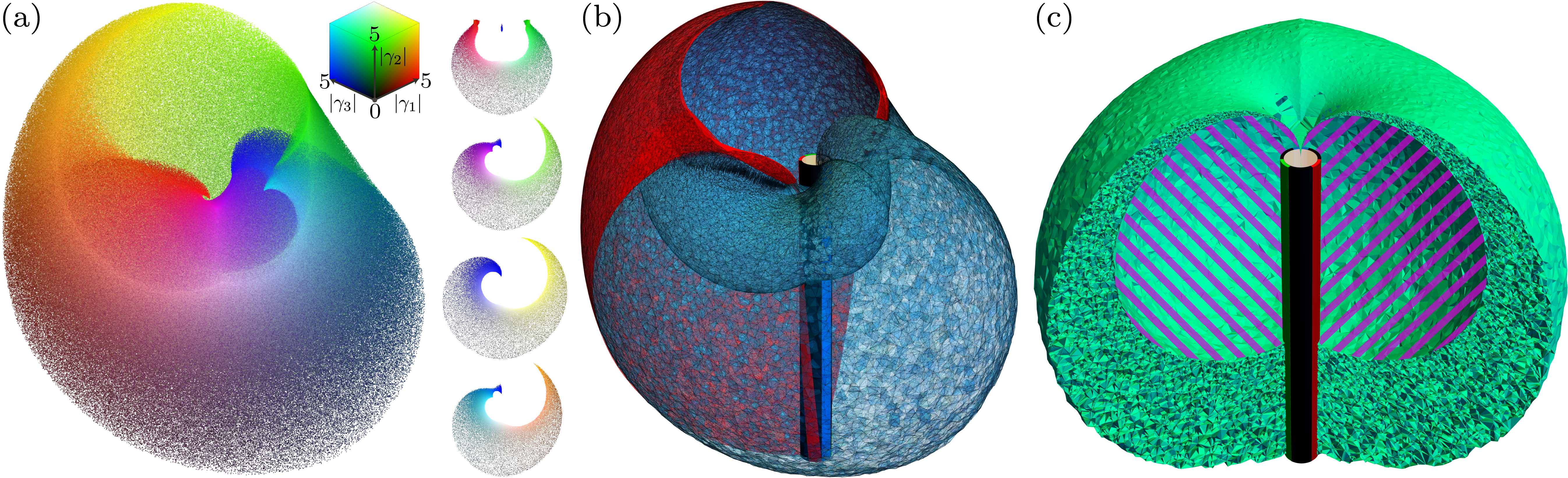}
\caption{{\bf{Reachability cloud for a minimal elephant trunk design with three actuators.}}
(a) The point cloud consists of 2 million points, RGB-coded by activation magnitude within the range $\gamma_1, \gamma_2, \gamma_3\in[-5,0]$ (red, green, and blue, respectively). (b) A design with only 2 helical actuators (shown in red) restricts the reachability cloud to a 2D surface region compared to 3D cloud of the minimal design (in blue). (c) Similarly, a design with 3 longitudinal actuators has a small reachability cloud with a large interior unreachable region.} 
\label{fig4}
\end{figure*}

\textit{Minimal elephant trunk-inspired design.}---Our study in Fig.~\ref{fig3} suggests that a single muscular bundle or two equal opposite helical bundles can generate a variety of deformation modes. By combining these modes, we can create a minimal design of a tubular structure made up of one longitudinal and two helical fibers. The longitudinal fiber generates pure curvature as inspired by the massive upper muscle group in the elephant trunk, see Fig.~\ref{fig1}. The two symmetrically arranged helical fibers generate either  twist, torsion, or curvature depending on their relative activation magnitudes; see Fig.~\ref{fig4} and the design parameters in the \smref{}.

To quantify the performance of this design, we created a comprehensive three-dimensional reachability point cloud in Fig.~\ref{fig4}a to illustrate the space of end points accessible by the trunk by randomly sampling  muscular activations. 
Notably, the reachability cloud consists of 2 million points and  takes less than 1 minute to compute on a standard desktop computer, at a rate of roughly 100,000 configurations per second thanks to the semi-analytical nature of the model and unlike traditional finite-element code.

The resulting cloud spans an extensive three-dimensional space within the entire $360^{\circ}$ arc around the longitudinal axis of the initial configuration. Our minimal trunk design can reach points to the front, back, right, and left of the initial configuration, as well as anywhere in between these regions. For comparison, the trunk of an Asian elephant can generate compressive strains up to 33\% \cite{wilson_continuum_1991}, which corresponds to a stretch of 0.67 and an activation of $\gamma \approx -3.33$ in our minimal design. Importantly, a tight concave hull of the reachability cloud with normalized volume $\tilde{V}_\text{cloud} = V_\text{cloud}/L^3 \approx 0.85$ amounts to about $74\%$ of the convex hull with $\tilde{V}_\text{conv} \approx 1.14$, which implies that only $26\%$ of the conservative volume bounded by the cloud is inaccessible by the design. Comparison with other designs containing either 2 helical actuators or 3 longitudinal ones shows in Fig.~\ref{fig4}bc that our minimal design is vastly superior in terms of reachability. Specifically, removing the blue longitudinal actuator reduces the reachability cloud to a 2D surface (Fig.~\ref{fig4}b), and replacing the helical actuators with longitudinal ones decreases the cloud volume by $14\%$ to $\tilde{V}_\text{cloud} \approx 0.73$ due to a large inaccessible void created by the removal of helical fibers, amounting to almost $42\%$ of the new convex hull (Fig.~\ref{fig4}c). 
Indeed, the torsional modes activated by the helical fibers empower the design to not only reach a larger space, but also perform a variety of complex motions, including avoiding obstacles or grasping and rotating objects through a curling motion, a common motion performed by elephant trunks when manipulating branches of trees. 

\begin{figure*}[t!]
\centering
\includegraphics[width=1\textwidth]{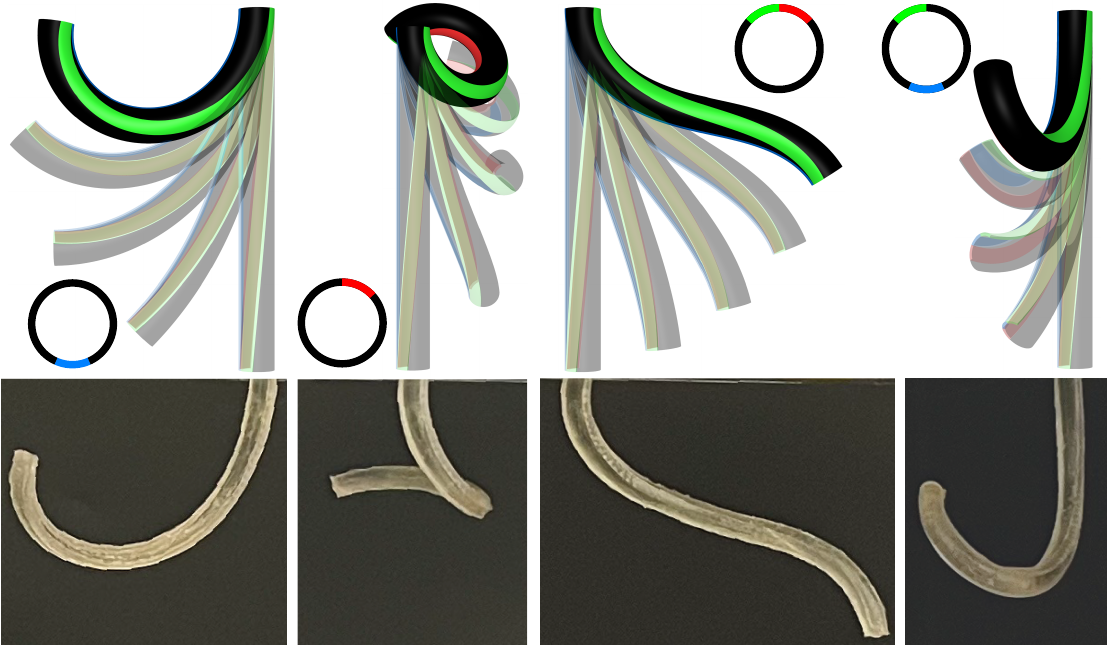}
\caption{{\bf{Experimental validation of predicted deformation patterns.}} 
Activation of the single longitudinal, a single helical, both helical, and a pair of longitudinal and helical fibers, from left to right.
The predicted deformation patterns, top, agree with the experimental deformation modes of a soft slender structure activated by liquid-crystal elastomer fibers, bottom.}
\label{fig5}
\end{figure*}
\textit{Experimental validation.}---To conclude our study, we validate our predicted deformation patterns with experimental deformation patterns of a soft slender structure actuated by liquid crystal elastomer fibers, which contract upon heating. Briefly, we created an inactive acrylate polymer-based slender cylinder via molding. Three active liquid crystal elastomer fibers were fabricated through 3D printing for mesogen alignment. Before curing, we embedded a coiled copper wire in each of the fibers to enable active contraction through Joule heating. We then cured the fibers with ultraviolet light to fix the mesogen alignment and lock the wire in place. Finally, we bonded the resulting fibers to the elastic cylinder, giving us the complete experimental model.

Fig.~\ref{fig5} shows a direct comparison of four representative modes from the activation of (i) the single longitudinal fiber, (ii) a single helical fiber, (iii) both helical fibers, and (iv) a pair of longitudinal and helical fibers, from left to right.
For a more accurate comparison with the experiment, the model predictions in Fig.~\ref{fig5} include the effects of gravity (see \smref{}). 
The four configurations in Fig.~\ref{fig5} illustrate the richness of possible deformation modes: configuration (i) exhibits {\it{pure bending}} achieved by activating the longitudinal actuator only; configuration (ii) exhibits high {\it{torsion}} through the activation of one of the helical actuators; configuration (iii) exhibits {\it{pure bending}} through the two equal and opposite helical actuators, with the bending direction changing halfway following the fiber arrangement; finally, configuration (iv) represents a {\it{combination of bending and torsion}}, generating an out-of-plane motion induced by the interplay of the longitudinal and helical actuators. 
The simulated configurations agree with the experimental deformation patterns, which supports the utility of our parametric studies and reachability considerations.

\textit{Conclusion.}---Unlike traditional robotic arms with finite degrees of freedom, soft structures like the elephant trunk or the octopus arm have infinitely many, hence creating ``a body of pure possibility" according to the philosopher Peter Godfrey \cite{godfrey2016other}. This universe of possibilities is made possible through fundamental physical mechanisms coupling geometry to mechanics, amplified by the {\it{slenderness}} of the structure and the {\it{softness}} of matter.  
While curvature generation in a one-dimensional structure is easy to understand and create through longitudinal contraction, the creation of twist and torsion requires dedicated helical fibers. To create both left- and right-handed structures, the design requires two helical bundles with opposite handedness. Hence, the simplest design is a three-actuator architecture, inspired by the elephant trunk. Here we demonstrate, analytically, computationally, and experimentally, that this minimal design is sufficient to reach a large portion of space. Further, the large range of deformation types empowered by this minimal design suggests that it could play an important role in 
motion planning tasks.\\[-4.pt]

\begin{acknowledgments}
\noindent \textbf{Funding}: This work has been funded by EPSRC grant EP R020205 and the NSF grant CMMI 2318188. \\
\textbf{Authors contribution}: BK, EK, DM, AG designed the study and developed the theory, 
BK performed the computations,
SL, RZ designed the actuators and performed the experiments. All authors analyzed the data, discussed the results, and wrote the paper.
\end{acknowledgments}

\bibliographystyle{apsrev4-2}
%
\end{document}


\title{Supplementary materials: Minimal design of the elephant trunk as an active filament}  
\author{Bartosz Kaczmarski, Sophie Leanza, Renee Zhao,}
\affiliation{Department of Mechanical Engineering,
Stanford University, Stanford, CA-94305, United States}
\author{Derek E.\ Moulton}
\affiliation{Mathematical Institute, University of Oxford, Oxford, UK}
\author{Ellen Kuhl}
\affiliation{Department of Mechanical Engineering,
Stanford University, Stanford, CA-94305, United States}
\author{Alain Goriely}
\affiliation{Mathematical Institute, University of Oxford, Oxford, UK}

\maketitle

\textit{Model set-up.}---In the active filament formulae,
\begin{equation}
\begin{split}
\hat{{\mathsf u}}_{1} &= -\frac{4}{3R_0^4}A\delta_1\sin\left(\varphi - \frac{Z}{R_0}\tan\alpha\right),\\
\hat{{\mathsf u}}_{2} &= -\frac{4}{3R_0^4}A\delta_2\cos\left(\varphi - \frac{Z}{R_0}\tan\alpha\right ),\\
\hat{{\mathsf u}}_{3} &= \frac{2}{R_0^4}\delta_3a_0,\quad \hat{\zeta} = 1 + \frac{1}{2R_0^2}\delta_0 a_0,
\end{split}
\label{eq:curvs}
\end{equation}
the quantities $\delta_0$, $\delta_1$, $\delta_2$, and $\delta_3$ are functions of the filament geometry, fiber architecture, and the Poisson's ratio of the filament material. As shown in \cite{kaczmarski2022active}, given a ring cross-section with an inner radius $R_1$ and an outer radius $R_2 = R_0$, a fiber architecture with a constant helical angle $\alpha$, and a Poisson's ratio $\nu$ of the structure, we have
\begin{widetext}
\begin{equation}
\begin{split}
\delta_{0}&=(\nu +1) R_2^2 \cot ^2\alpha  \log
   \left(\frac{R_2^2 \sec ^2\!\alpha }{R_1^2 \tan
   ^2\alpha +R_2^2}\right)+\nu 
   \left(R_1^2-R_2^2\right), \\
\delta_{1}&=\delta_{2}=\left(R_1-R_2\right) \left(\nu  \left(R_1^2+R_2
   R_1+R_2^2\right)-3 (\nu +1) R_2^2 \cot ^2\alpha\right)\\
   &\phantom{\delta_{2}=\delta_{1}=}+3 (\nu +1) R_2^3 \cot ^3\alpha 
   \arctan\left(\frac{R_1 \tan \alpha}{R_2}\right)-3 \alpha (\nu +1) R_2^3 \cot ^3\alpha,\\
\delta_{3}&=- (1+\nu) R_2 \cot\alpha 
   \left(R_1^2-R_2^2+R_2^2 \cot ^2\alpha
   \log \left( \frac{2 R_2^2}{\left(R_2^2-R_1^2\right) \cos 2
   \alpha+R_1^2+R_2^2}\right)
   \right).
\end{split}
\label{eq:deltas1}
\end{equation}
\end{widetext}
For longitudinal fibers, we  take the limit $\alpha\rightarrow 0$, for which these expressions simplify to
\begin{equation}
\begin{split}
\delta_{0}&=R_2^2 - R_1^2, \\
\delta_{1}&=\delta_{2}=R_2^3 - R_1^3,\\
\delta_{3}&=0.
\end{split}
\end{equation}
In the case of a tapered filament geometry with a tapering angle $\phi_2(Z)$ defined at the outer surface $R = R_2(Z)$ of the filament, and longitudinal fibers with $\alpha = 0$, the $\delta_0$, $\delta_1$, $\delta_2$, and $\delta_3$ quantities then become
\begin{equation}
\begin{split}
\delta_{0}&=4(R_1^2-R_2^2)\nu-\frac{4(1+\nu)}{c_\phi^2}\log\left(\frac{1 + R_1^2 c_\phi^2}{1 + R_2^2 c_\phi^2}\right), \\
\delta_{1}&=\delta_{2}=4(R_1^3-R_2^3)\nu+\frac{12(1+\nu)}{c_\phi^3}\left(\arctan(R_1 c_\phi) \right.\\&\qquad\qquad\left.- \arctan(R_2 c_\phi) + (R_2 - R_1)c_\phi\right),\\
\delta_{3}&=0,\\
c_\phi &= c_\phi(Z) = \frac{\tan{(\phi_2(Z))}}{R_2(Z)}.
\end{split}
\label{eq:deltasT}
\end{equation}
\textit{Curvatures in the parametric studies}---In the parametric studies, we reported the dimensionless curvature $L\kappa$ for the case of varied slenderness $S_\text{r}$, and the maximal curvature $L\kappa_\text{max}$ for the cases in which the tapering angle $\phi$ and the fiber revolution $\Omega$ were varied. The reason for reporting the maximal curvature in the two latter scenarios is that the curvature itself varies along the length of the filament. To supplement the reported results, fig.\ S1 provides specific numerical values of $L\kappa$ and $L\kappa_\text{max}$ for the 12 evaluated designs, as well as the curvature functions $L\kappa(Z)$ for the tapering angle and fiber helicity studies shown in figs.\ S1b and S1c, respectively.
\begin{figure*}[t!]
\centering
\includegraphics[width=1\textwidth]{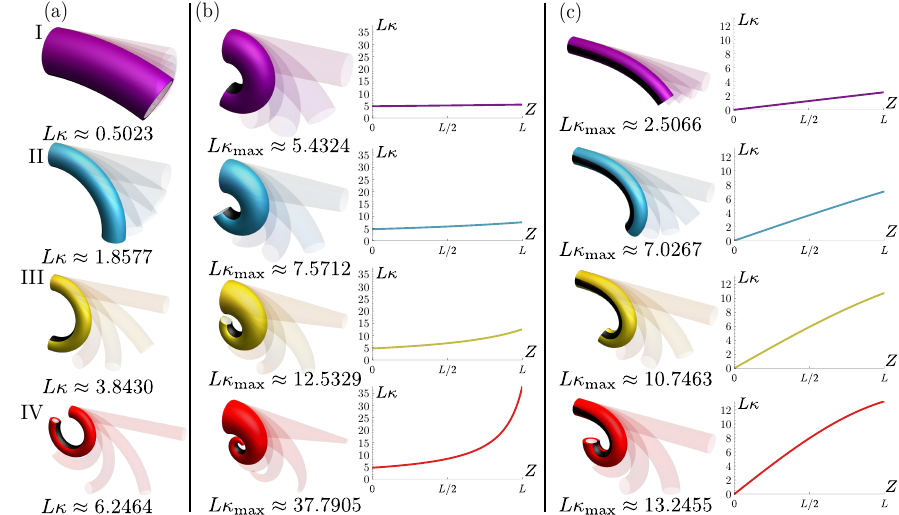}
\caption{Additional information regarding curvature values and curvature functions present for the 12 designs presented in the parametric studies. (a) Designs with different slenderness values. Since the curvature is constant along $Z$, just the values of $L\kappa$ are shown for the designs I-IV. (b) Designs with different tapering angles and (c) designs with different fiber revolution angles. The curvature varies along $Z$ in (b) and (c), so both the values of $L\kappa_\text{max}$ and the plots of $L\kappa$ vs.\ Z are shown for the four respective designs in each case.}
\label{FIGS1}
\end{figure*}

\textit{Material properties of evaluated designs.}---In  parametric studies, the only mechanical property of the filament material whose value is significant for the reported results is the Poisson's ratio $\nu$, which was chosen to be $\nu = 1/2$ and homogeneous for all 12 designs. The Young's modulus $E$ was also assumed to be homogeneous throughout the entire structure in those designs, which, according to the theory in \cite{kaczmarski2022active}, removes the dependence on the value of $E$ in $\hat{{\mathsf u}}_{1}$, $\hat{{\mathsf u}}_{2}$, $\hat{{\mathsf u}}_{3}$, and $\hat{\zeta}$ under the assumption of the ring solution.

However, for the comparison between the minimal design and the experimental results to be valid, the assumption of a homogeneous Young's modulus and Poisson's ratio is no longer valid since the LCE fiber bundles in the experimental prototype have  different material properties than that of the surrounding elastic matrix.

Therefore, a minor adjustment to the model setup has to be made to account for  material inhomogeneity in the minimal design. Let $E_\text{f}$ and $\nu_\text{f}$ denote the elastic modulus and Poisson's ratio of the LCE fibers, and $E_\text{M}$ and $\nu_\text{M}$ denote the elastic modulus and Poisson's ratio of the elastic matrix. Then, an approximation of the intrinsic curvatures $\hat{{\mathsf u}}_{1}$, $\hat{{\mathsf u}}_{2}$, $\hat{{\mathsf u}}_{3}$, and extension $\hat{\zeta}$ can be shown to be expressed as
\begin{widetext}
\begin{equation}
\begin{split}
\hat{{\mathsf u}}_{1} &\approx -\frac{4E_\text{f}}{3 \left(E_\text{M}R_1^4 + E_\text{f}(R_0^4 - R_1^4)\right)} A\delta_{1,\text{f}}\sin \left(\varphi - \frac{Z}{R_0}\tan\alpha\right),\\
\hat{{\mathsf u}}_{2} &\approx  -\frac{4E_\text{f}}{3 \left(E_\text{M}R_1^4 + E_\text{f}(R_0^4 - R_1^4)\right)} A\delta_{2,\text{f}}\cos \left(\varphi - \frac{Z}{R_0}\tan\alpha\right),\\
\hat{{\mathsf u}}_{3} &\approx  \frac{2E_\text{f}(1 + \nu_\text{M})}{E_\text{M}R_1^4(1 + \nu_\text{f}) + E_\text{f}(R_0^4 - R_1^4)(1+\nu_\text{M})} \delta_{3,\text{f}} a_0,\\
\hat{\zeta} &\approx  1 + \frac{E_\text{f}}{2(E_\text{M}R_1^2 + E_\text{f}(R_0^2-R_1^2))}\delta_{0,\text{f}} a_0,
\end{split}
\label{eq:inhomogeneous1}
\end{equation}
\end{widetext}
where $\delta_{i,\text{f}}$ are defined as in Eqs.\ \eqref{eq:deltas1} and \eqref{eq:deltasT}, but with the substitution $\nu \rightarrow \nu_\text{f}$. All other quantities are defined as before. The approximation stems from the simplifying assumption that the activated ring region shares the material properties of the LCE fibers, even if the LCE fibers do not make up the entirety of the activatable ring. The accuracy of this approximation increases as the size of the ring relative to the size of the elastic matrix decreases, and as the total angular extent of the LCE fibers in the ring approaches $2\pi$.

\textit{Multiple helical angles in a ring.}---The minimal design consists of a longitudinal fiber bundle and two symmetrically arranged helical fiber bundles of opposite handedness. That is, $\alpha^{(1)} = 0 \neq \alpha^{(2)} \neq \alpha^{(3)}$ and $\alpha^{(2)} = -\alpha^{(3)}$, where $\alpha^{(1)} = 0$ is the helical angle of the longitudinal fiber bundle, and $\alpha^{(2)}$, $\alpha^{(3)}$ are the helical angles of the two helical fiber bundles. However, Eq.\ \eqref{eq:curvs} is posed in terms of a single angle $\alpha$. To be able to account for the contributions of all three fiber bundles, we can utilize the superposition property of the active filament model, which handles the mathematical treatment of multiple concentric rings of activation \cite{kaczmarski2022active}. Similar to Eq.\ \eqref{eq:inhomogeneous1}, for a general filament with $M$ concentric rings of activation and different material properties in the fibers of each ring and the elastic matrix, the resulting curvatures and extensions can be approximated as
\begin{widetext}
\begin{equation}
\begin{split}
\hat{{\mathsf u}}_{1} &\approx
-\frac{4\sum_{i=1}^M E_\text{f}^{(i)}\delta_{1,\text{f}}^{(i)}A^{(i)}\sin\left(\varphi^{(i)} - \frac{Z}{R_2^{(i)}}\tan{\alpha^{(i)}}\right)}{3\left[E_\text{M}\mleft(R_1^{(1)}\mright)^4 + \sum_{i = 1}^M E_\text{f}^{(i)}\left(\mleft(R_2^{(i)}\mright)^4-\mleft(R_1^{(i)}\mright)^4\right)\right]},\\
\hat{{\mathsf u}}_{2} &\approx
-\frac{4\sum_{i=1}^M E_\text{f}^{(i)}\delta_{2,\text{f}}^{(i)}A^{(i)}\cos\left(\varphi^{(i)} - \frac{Z}{R_2^{(i)}}\tan{\alpha^{(i)}}\right)}{3\left[E_\text{M}\mleft(R_1^{(1)}\mright)^4 + \sum_{i = 1}^M E_\text{f}^{(i)}\left(\mleft(R_2^{(i)}\mright)^4-\mleft(R_1^{(i)}\mright)^4\right)\right]},\\
\hat{{\mathsf u}}_{3} &\approx
\frac{2\sum_{i=1}^M a_0^{(i)}E_\text{f}^{(i)}\delta_{3,\text{f}}^{(i)}\left(1 +\nu_\text{f}^{(i)}\right)^{-1}}{E_\text{M}\mleft(R_1^{(i)}\mright)^4\left(1 + \nu_\text{M}\right)^{-1} + \sum_{i = 1}^M E_\text{f}^{(i)}\left(\mleft(R_2^{(i)}\mright)^4 - \mleft(R_1^{(i)}\mright)^4\right)\left(1 + \nu_\text{f}^{(i)}\right)^{-1}},\\
\hat{\zeta} &\approx 1 + \frac{\sum_{i = 1}^M a_0^{(i)} E_\text{f}^{(i)} \delta_{0,\text{f}}^{(i)}}{2\left[E_\text{M}\mleft(R_1^{(i)}\mright)^2 + \sum_{i = 1}^M E_\text{f}^{(i)}\left(\mleft(R_2^{(i)}\mright)^2-\mleft(R_1^{(i)}\mright)^2\right)\right]},
\end{split}
\label{eq:multiring}
\end{equation}
\end{widetext}
where any quantity $(\cdot)^{(i)}$ is associated with the $i$-th activation ring in the filament, $i = 1,\ldots, M$. The approximation is again the result of the assumption of constant material properties $E_\text{f}^{(i)}$, $\nu_\text{f}^{(i)}$ within a given ring, but different properties $E_\text{M}$, $\nu_\text{M}$ in the elastic matrix. We emphasize that, in the single-ring case, it was true that $R_0 = R_2$ (outer radius of the whole structure is the outer radius of the single ring), but, in the $M$-ring case, we have $R_0 = R_2^{(M)}$.

In the case of the minimal design, three rings are needed to represent the actuators utilizing the three different fiber architectures. All three rings have the same inner radius $R_1$ and outer radius $R_2$ since all three fiber bundles are contained within one ring region. Therefore, we can set $R_1^{(1)} = R_1^{(2)} = R_1^{(3)} = R_1$ and $R_2^{(1)} = R_2^{(2)} = R_2^{(3)} = R_0$. Further, the three fiber bundles have the same material properties, so $E_\text{f}^{(1)} = E_\text{f}^{(2)} = E_\text{f}^{(3)} = E_\text{f}$ and $\nu_\text{f}^{(1)} = \nu_\text{f}^{(2)} = \nu_\text{f}^{(3)} = \nu_\text{f}$, which simplifies the general expressions in Eq.\ \eqref{eq:multiring}.

The outlined procedure was also applied to the simpler case of the design with two symmetrically arranged helical fiber bundles in the parametric study of the fiber revolution parameter $\Omega$. In that scenario, we remove the longitudinal actuator ring and assume $E_\text{M} = E_\text{f}^{(i)}$, $\nu_\text{M} = \nu_\text{f}^{(i)}$, for $i = 1,2$.

\textit{Definitions of tapering and fiber revolution}---For a general tapering profile, we can define the tapering angle field as a function of the cylindrical radius coordinate $R$ and the parameter $Z$ as \cite{kaczmarski2022active}
\begin{align}
\tan(\phi(R, Z)) = \frac{R}{R_2(Z)}R_2'(Z).
\end{align}
Given that the tapering profile is assumed to be linear, it can be described with a single tapering angle prescribed for the whole design such that
\begin{align}
\tan(\phi(R=R_2)) = \frac{R_2(L) - R_2(0)}{L},
\end{align}
which is the definition used for the color bar in the parametric studies. The inner radius profile $R_1(Z)$ is then set to follow the tapering profile according to
\begin{align}
R_1(Z) = \frac{R_1(0)}{R_2(0)}R_2(Z).
\end{align}

The fiber revolution angle $\Omega$ is the total angular revolution of the fiber for $Z\in[0,L]$. Given a helical angle $\alpha$, it is  defined for an non-tapered structure as
\begin{align}
\Omega = \frac{L}{R_2}\tan{\alpha}.
\end{align}
Using the fiber revolution angle for design purposes is the most intuitive as it allows direct reasoning about fiber interpenetration when multiple helical fibers are present.

\textit{Stiffness correction for overlapping rings}---The actuators of the minimal design are represented with three activation rings with the same inner and outer radii. Thus, the rings overlap fully with one another, which results in overcounting of the stiffness contributions from $E_\text{f}$ in each of the overlapping rings, while there is only one physical ring region. As such, we use a simple stiffness correction that addresses this overcounting issue. Since the stiffness is overcounted $M$ times for an $M$-ring filament, a simple scaling by $k_\text{count} = 1/M$ is sufficient, so that the corrected Young's modulus is $\tilde{E}_\text{count}^{(i)} = k_\text{count} E_\text{f}$ for the $i$-th ring.

\textit{Stiffness correction for the comparison with experimental results}---The fabricated experimental prototype involves LCE fiber bundles that adhere to the outer surface of the tubular elastic matrix. In the context of the ring solution in the active filament model, this effectively introduces empty space between the activated angular sectors of the ring. By default, however, the model assumes that the elastic matrix also fills the space between the fiber bundles within the ring, since the derivation of the curvature and extension formulae is not straightforward for the case of a Young's modulus that is variable and discontinuous in the polar angle. Thus, we performed a stiffness averaging operation, so that the adjusted axial stiffness matches the axial stiffness of the ring with the empty space that is present in the experimental setup.

Specifically, consider the $i$-th ring of the filament with a Young's modulus $E_\text{f}$ and with its default definition under the active filament model (filled space between fiber bundles). Its axial stiffness is 
\begin{align}
    K_0^{(i)} = E_\text{f}\pi((R_2^{(i)})^2 - (R_1^{(i)})^2).
\end{align}
Now, for a piecewise-constant activation pattern $\gamma^{(i)}(\theta)$ that defines a set of $N^{(i)}$ discrete fiber bundles with angular extents $\sigma^{(i)}$ in the $i$-th ring, the axial stiffness of the ring that accounts for the empty space is
\begin{align}
    \tilde{K}_0^{(i)} = \frac{N^{(i)} \sigma^{(i)}}{2\pi} K_0^{(i)},
\end{align}
which means that the scaling factor for the Young's modulus of the $i$-th ring could be chosen as $k_\text{empty}^{(i)} = N^{(i)} \sigma^{(i)}/2\pi$, so that the corrected Young's modulus is $\tilde{E}_\text{empty}^{(i)} = k_\text{empty}^{(i)}E_\text{f}$.

We combine the stiffness correction for overlapping rings with the empty-space stiffness correction, which gives the final corrected Young's modulus
\begin{align}
\tilde{E}_\text{f} = \mleft(k_\text{count}\sum_{i = 1}^M k_\text{empty}^{(i)}\mright)E_\text{f}.
\end{align}
For consistency, this combined correction was also employed in the generation of the reachability clouds in addition to its use in the comparison with experimental deformations, given that the same actuator design is used in both results. 

Note that the discussed stiffness corrections are not relevant to the parametric study of the fiber revolution angle, since that scenario assumes a homogeneous Young's modulus throughout the filament, i.e., $E_\text{M} = E_\text{f}$, which removes the dependence of the curvature and extension functions on the Young's modulus. Particularly, the stiffness overcounting is not manifested in that case, even though two overlapping helical rings are used. There is also no need to apply the empty-space correction, as the default active filament representation is appropriate for reasoning about the fiber helicity effects.

\textit{Minimal design parameter details}---The minimal design consists of a filament of length $L = 0.09 \,\text{m}$ and an outer radius $R_0 = L/20$, in which the activatable ring has a constant outer radius $R_2 = R_0 = L / 20$ and a constant inner radius $R_1 = (5/6)R_2$. The piecewise-constant activation pattern in the ring defines a set of three fiber bundles with angular extents $\sigma = 48^{\circ}$. Two of the fiber bundles are helical with opposite handedness (red and green actuators), and start with angular phase offsets $\theta_{0,\text{red}} = 66^{\circ}$ and $\theta_{0,\text{green}} = 114^{\circ}$ in the cross section at $Z = 0$. The third fiber bundle (blue actuator) is longitudinal and starts at an angle offset $\theta_{0,\text{blue}} = 270^{\circ}$ in the cross section. The fiber revolutions of the helical fiber bundles are set such that they just meet (and not interpenetrate) the longitudinal fiber bundle at $Z = L$, which means that $\Omega_\text{red} = -108^{\circ}$ and $\Omega_\text{green} = 108^{\circ}$. These give rise to helical angles $\alpha_\text{red} \approx -0.756$ and $\alpha_\text{green} \approx 0.756$.

The material properties of the minimal design are set to match the experimental setup in the experimental validation step. In particular, we set
\begin{equation}
\begin{split}
E_\text{f} &= 1.4 \, \text{MPa},\\
\nu_\text{f}&= 0.495,\\
E_\text{M} &= 127.8 \,\text{kPa},\\
\nu_\text{M} &= 0.5,
\end{split}
\label{eq:mindesignprop}
\end{equation}
and a volumetric density $\rho_\text{vol} = 1000 \text{kg}/\text{m}^3$.

\begin{figure}[t!]
\centering
\includegraphics[width=0.82\columnwidth]{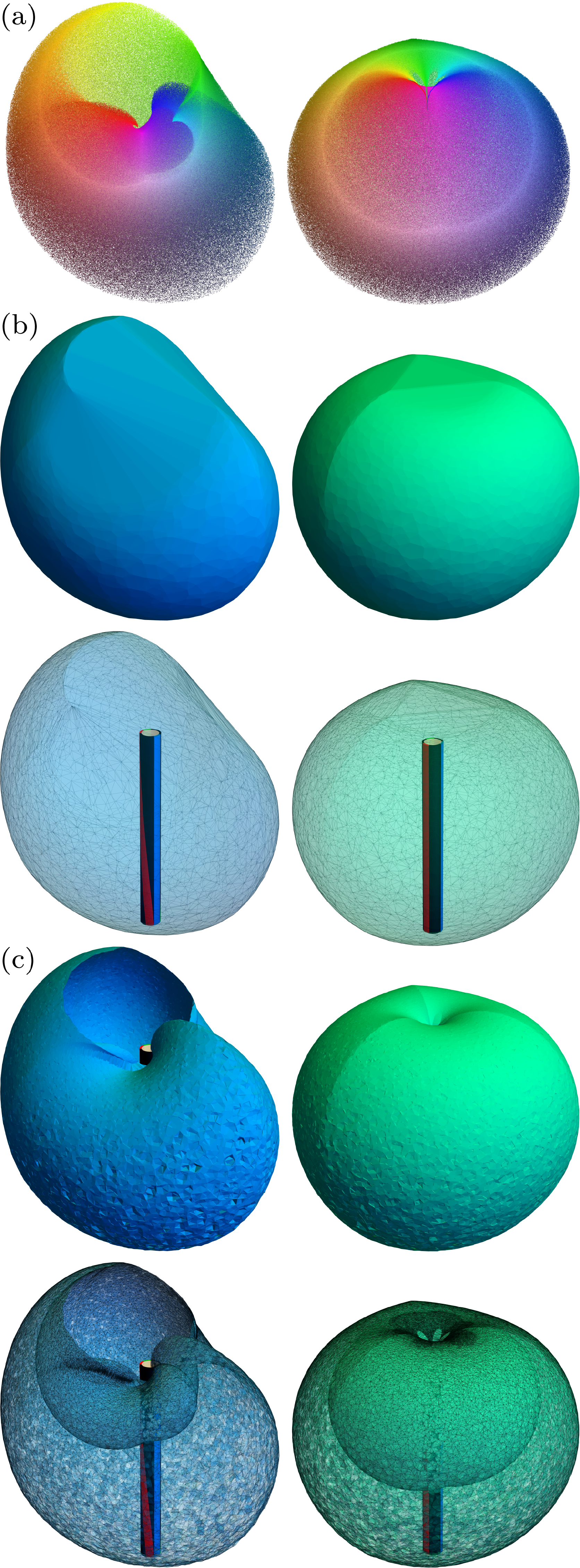}
\caption{Convex and concave hulls of the reachability clouds of the minimal design (left) and the design with 3 longitudinal fiber bundles (right). (a) The discrete reachability point cloud of deformed design endpoints obtained through uniform sampling $\gamma_i\in[-5,0]$. (b) The convex hull of the cloud which gives the upper bound for the cloud volume. (c) Concave hull mesh that wraps around the topologically complex, non-convex structure of the cloud. The initial configurations are shown in the translucent visualizations of the hull meshes.}
\label{FIGS2}
\end{figure}

\textit{Reachability cloud volume}---The task of establishing the volumetric coverage of an arbitrary point cloud is generally non-trivial. We estimated the upper bound of the reachability cloud volume as the volume of the convex hull bounding the 2 million endpoints of the deformed design. Further, we estimated the minimal non-convex volume spanned by the reachability cloud by constructing the concave hull mesh of the 3D data using the $\alpha$-shapes method \cite{edelsbrunnerShapeSetPoints1983}. Both the convex and the concave hulls are visualized in Fig.\ S\ref{FIGS2} for both the minimal design and the design with 3 longitudinal fiber bundles.

\textit{Muscle strains in the minimal design}---To compare the activation magnitudes $|\gamma_i|$ to muscular contraction levels observed physiologically in elephant trunks, we used a measure of deformed fiber strain during a simple in-plane bending motion of the minimal design with only the longitudinal bundle activated. An estimate of the deformed fiber strain in the deformed configuration can be computed as $(L_\mathcal{F}^1/L_\mathcal{F}^0) - 1$, where $L_\mathcal{F}^0$ is the initial total arc length of the fiber $\mathcal{F}$ and $L_\mathcal{F}^1$ is its deformed total arc length. For a given filament configuration $\{\mathbf{r}(Z), \mathbf{d}_1(Z), \mathbf{d}_2(Z), \mathbf{d}_3(Z)\}$, the total arc length of a fiber with initial phase offset $\theta_0$, helical angle $\alpha$, and outer radius $R_2$ can be computed as
\begin{align}
L_\mathcal{F} &= \int_0^L \left\|\frac{\partial}{\partial Z}\left[\mathbf{r}(Z) + R_2\left(\mathbf{d}_1\cos\left(\theta_0 + \frac{Z}{R_2}\tan\alpha\right)\right.\right.\right. \nonumber\\
&\hspace{1.5cm}\left.\left.\left.+ \;\mathbf{d}_2\sin\left(\theta_0 + \frac{Z}{R_2}\tan\alpha\right)\right)\right]\right\|\,dZ.
\end{align}
For the minimal design, this gives an estimate of $|\gamma_3|\approx 3.33$ that results in a fibrillar contraction of $33\%$ in the longitudinal bundle for the pure-bending deformation.

\textit{Filament integration with gravitational forces}---To compute the deformation of the activated filament under the influence of gravitational forces due to its own weight, we treat the filament as an extensible, unshearable Kirchhoff rod governed by a constitutive law for an isotropic material with a quadratic energy form. This construction is adapted from \cite{molego12} for the specific case of integration under self-weight.

Let us first specify three configurations utilized in the theory: the initial configuration (pre-activation), the reference configuration (post-activation) and the deformed configuration (post-activation and with external loading). The arc length parameters for the three configurations are $Z$, $S$, and $s$, respectively. The extension in the reference configuration is $\hat{\zeta} = dS/dZ$, and the extension in the deformed configuration due to external forces is $\zeta_\mathbf{F} = ds/dS$.

Under the influence of gravitational acceleration $\mathbf{g}$, the filament is subject to an external body force per unit length in the deformed configuration $\rho\mathbf{g}$, where $\rho$ is the linear density of the filament in the deformed configuration. This gives
\begin{align}
\frac{\partial \mathbf{n}}{\partial s} +\rho \mathbf{g} &=\mathbf{0}
\end{align}
for the balance of the internal force $\mathbf{n}$ in the deformed configuration. Now, given that $ds/dZ = (ds/dS)(dS/dZ)=\zeta_\mathbf{F}\hat{\zeta}$, we can write
\begin{align}
\frac{\partial \mathbf{n}}{\partial Z} +\hat{\zeta}\zeta_\mathbf{F}\rho\mathbf{g} &=\mathbf{0}.
\end{align}
Further, an infinitesimal element of mass $dm$ and density $\rho_0$ in the initial configuration is mapped to an element with the same mass and different density $\rho$ in the deformed configuration, such that
\begin{align}
dm = \rho\,ds = \rho\hat{\zeta}\zeta_\mathbf{F}\,dZ = \rho_0\, dZ \Rightarrow \rho = \frac{\rho_0}{\hat{\zeta}\zeta_{\mathbf{F}}} 
\end{align}
Thus,
\begin{align}
\frac{\partial \mathbf{n}}{\partial Z} +\rho_0 \mathbf{g} &=\mathbf{0},
\end{align}
which, given that $\mathbf{n}(L) = \mathbf{0}$, yields the internal force
\begin{align}
    \mathbf{n}(Z) = \mathbf{g}\int_{Z}^L \rho_0(Z')\, dZ'.
\end{align}
Assuming a non-tapered filament with homogeneous volumetric density ($\rho_0$ becomes constant) and that $\mathbf{g} = -g\mathbf{e}_Z$, we obtain the following components of $\mathbf{n}$ expressed in the local basis:
\begin{align}
    \mathsf{n}_i(Z) = g\rho_0(Z-L)(\mathbf{d}_i\cdot\mathbf{e}_Z).
\end{align}

The general moment balance for the filament can be expressed in the deformed configuration as
\begin{align}
\frac{\partial \mathbf{m}}{\partial s} +\frac{\partial \mathbf{r}}{\partial s}\times \mathbf{n} +\zeta_\mathbf{F}^{-1}\mathbf{l} &=\mathbf{0},
\end{align}
where $\mathbf{m}$ is the internal moment, and $\zeta_\mathbf{F}^{-1}\mathbf{l}$ is the couple per unit length in the deformed configuration. For gravitational loading, we have $\mathbf{l} = \mathbf{0}$, so we can write the components of the moment $\mathbf{m}$ in the local basis as
\begin{equation}
\begin{split}
\mathsf{m}_1'(Z) &= \hat{\zeta}(\mathsf{u}_3(Z) \mathsf{m}_2(Z) - \mathsf{u}_2(Z)\mathsf{m}_3(Z)) + \zeta\mathsf{n}_2(Z),\\
\mathsf{m}_2'(Z) &= \hat{\zeta}(\mathsf{u}_1(Z) \mathsf{m}_3(Z) - \mathsf{u}_3(Z)\mathsf{m}_1(Z)) - \zeta\mathsf{n}_1(Z),\\
\mathsf{m}_3'(Z) &= \hat{\zeta}(\mathsf{u}_2(Z) \mathsf{m}_1(Z) - \mathsf{u}_1(Z)\mathsf{m}_2(Z)),
\end{split}
\end{equation}
where $\zeta = \zeta_\mathbf{F} \hat{\zeta}$ is the overall extension of the rod. The deformed curvatures $\mathsf{u}_i$ are derived from the intrinsic curvatures as
\begin{align}
\mathsf{u}_i(Z) &= \hat{\mathsf{u}}_i(Z) + \frac{\mathsf{m}_i(Z)}{K_i},  
\end{align}
where $K_i$ are the stiffness coefficients of the rod about $\mathbf{d}_i$, for $i = 1,2,3$.

To construct the extension $\zeta_\mathbf{F}$ due to the external loading, we use the constitutive relationship
\begin{equation}
\begin{split}
\mathsf{n}_3(Z) = K_0(\zeta_\mathbf{F}(Z) - 1).
\end{split}
\end{equation}

Finally, the deformed configuration of the rod is obtained by solving the boundary value problem
\begin{equation}
\begin{split}
\mathbf{r}'(Z) &= \zeta(Z) \mathbf{d}_3(Z),\\
\mathbf{d}_i'(Z) &= \hat{\zeta}(Z) \mathbf{u}(Z)\times \mathbf{d}_i(Z), \quad i\in\{1,2,3\},\\
\mathsf{m}_1'(Z) &= \hat{\zeta}(\mathsf{u}_3(Z) \mathsf{m}_2(Z) - \mathsf{u}_2(Z)\mathsf{m}_3(Z)) + \zeta\mathsf{n}_2(Z),\\
\mathsf{m}_2'(Z) &= \hat{\zeta}(\mathsf{u}_1(Z) \mathsf{m}_3(Z) - \mathsf{u}_3(Z)\mathsf{m}_1(Z)) - \zeta\mathsf{n}_1(Z),\\
\mathsf{m}_3'(Z) &= \hat{\zeta}(\mathsf{u}_2(Z) \mathsf{m}_1(Z) - \mathsf{u}_1(Z)\mathsf{m}_2(Z)),\\
\mathbf{r}(0) &= \mathbf{r}_0,\\
\mathbf{d}_i(0) &= \mathbf{d}_{i0}, \quad i\in\{1,2,3\},\\
\mathbf{m}(L) &= \mathbf{0},
\end{split}
\end{equation}
where $\mathbf{r}_0$, and $\mathbf{d}_{i0}$ define the clamped boundary condition. The BVP is solved using a shooting method and the Verner's 7(6) Runge-Kutta method for the integration step \cite{vernerNumericallyOptimalRunge2010}. Continuation is used through iterative scaling of the gravitational acceleration $g$ and reusing the previous BVP solution at each iteration.

\textit{Experimental methods}---To fabricate the LCE ink, liquid crystal mesogens (2-Methyl-1,4-phenylene bis(4-((6-(acryloyloxy)hexyl)oxy)benzoate)) (RM 82, BOC Sciences, USA) and (1,4-Bis-[4-(3-acryloyloxypropyloxy)benzoyloxy]-2-methylbenzene) (RM 257, BOC Sciences, USA) are mixed in a weight ratio of 2.7:1. Next, the spacers, (2,2-(Ethylenedioxy)diethanethiol) (EDDET, Sigma Aldrich, USA) and (2,6-Di-tert-butyl-4-methylphenol) (BHT, Sigma Aldrich, USA) are added at 22.5 wt\% and 2.25 wt\% to the total weight of mesogens. After melting all components at 80°C for 45 min, the catalyst triethylamine (TEA, Sigma Aldrich, USA) and photoinitiator Irgacure 819 (Sigma Aldrich, USA) are added at 1.6 wt\% and 1.8 wt\%, respectively, to the total weight of mesogens. The mixture is stirred for 3 min with a magnetic stir bar at 80°C, then oligomerized in an oven at 80°C for 15 min, and next transferred to a 10 mL syringe barrel (Nordson EFD, USA), and heated again in a vacuum oven at 80°C for 20 min. The ink is finally defoamed in a planetary mixer (AR-100, Thinky, USA) at 2200 rpm for 6 min to remove trapped air. A customized three-dimensional (3D) printer is used for printing of the LCE fibers. 90 mm-long fibers with longitudinal alignment, a width of 1.7 mm, and thickness of 0.4 mm are printed. 44 AWG (0.050 mm diameter) copper wire is coiled, with coil diameter of 0.5 mm, and stretched such that there are 0.68 coils/mm along the fiber. Two rows of coiled copper wires are embedded in each LCE fiber, and the fibers are then cured with UV light for 10 min. The Young's modulus of the cured LCE is measured to be 1.4 MPa.

The liquid resin for the inactive polymer rod is created by mixing butyl acrylate (Sigma Aldrich, USA), a commercial UV-curable phenoxyethyl acrylate monomer (Ebecryl 114, Allnex, USA), and a commercial UV-curable urethane oligomer (Ebecryl 8413, Allnex, USA) in a weight ratio of 12:7:1. Photoinitiator Irgacure 819 (Sigma Aldrich, USA) is added at 2 wt\% to the total weight of the resin, and the resin is mixed for 10 min at 80°C until homogeneous. The shear modulus of this inactive polymer upon curing is estimated as 42.6 kPa, using a neo-Hookean fit of tensile test data.

Silicone molds are created in order to fabricate the soft actuator. The mold is an inverse of the soft actuator design; it includes the cavity of a 4 mm diameter by 90 mm long cylinder with 0.4 mm thick slots along the cylinder, two patterned helically with opposite handedness and one patterned longitudinally. By inserting the fabricated LCE fibers into the slots in the mold, the orientation of the LCE fibers is fixed. The inactive polymer resin is then poured into the mold and UV light is shined on the semi-transparent mold for 15 minutes. During this time, the inactive rod cures while also bonding to the LCE fibers. Once removed from the mold, the soft actuator is fully fabricated. Currents between 0.9A and 1A are applied to the copper wires in the respective LCE fibers for the experimental implementations shown in Fig.\ 5.
\bibliographystyle{apsrev4-2}
\bibliography{elephant-trunk-SM}